\documentclass[12pt]{article}
\usepackage{graphicx}
\usepackage{mathrsfs}

\newcommand{\bfb}[1]{
            \mbox{\boldmath $ #1 $}}
\begin{document}

    \title{\bf{Probing Quantum Nonlinearities through Neutrino Oscillations}}

    \author{Wei Khim Ng\footnote{Email: phynwk@nus.edu.sg } \ and Rajesh R. Parwani\footnote{Email: parwani@nus.edu.sg}}


\maketitle

\begin{center}
{Department of Physics,\\}
{National University of Singapore,\\}
{Kent Ridge,\\}
{ Singapore.}

\end{center}

\begin{abstract}

We investigate potential quantum nonlinear corrections to Dirac's equation through its sub-leading effect on neutrino oscillation probabilities. Working in the plane-wave approximation and in the $\mu-\tau$ sector, we explore various classes of nonlinearities, with or without an accompanying Lorentz violation. The parameters in our models are first delimited by current experimental data before they are used to estimate corrections to oscillation probabilities. We find that only a small subset of the considered nonlinearities have the potential to be relevant at higher energies and thus possibly detectable in future experiments. A falsifiable prediction of our models is an energy dependent effective mass-squared, generically involving fractional powers of the energy.

\end{abstract}


\section{Introduction}
It appears that constant neutrino masses, though still not directly confirmed, are the simplest way of
explaining current data on neutrino oscillations \cite{review}. Other possibilities, such as Lorentz violating
dispersion relations \cite{coleman,coleman2}, do not seem to be possible explanations of the leading order
effects \cite{sub,sub1}.

Neutrinos are a valuable probe of new physics because of their weak interactions and in this paper we will study how a nonlinear modification to the quantum mechanical propagation of a neutrino, that is a nonlinear Dirac equation, affects neutrino oscillations. The propagating neutrino wavefunctions are assumed to follow
 \begin{equation}
        \left(i\gamma^\mu\partial_\mu-m+F\right)\psi=0 \,  , \label{nld}
\end{equation}
where $F$ is a matrix in spinor space, depending on the wavefunction $\psi$, its adjoint and their derivatives
\cite{NP1,np1a}. One may think of $F$ as probing deviations from exact quantum-linearity, a possibility that had
some empirical tests in the non-relativistic regime placing bounds on the nonlinearity scale
\cite{nonlin1,nonlin1a,nonlin1b,nonlin2,nonlin2a,nonlin2b,nonlin2c}.

In \cite{RP1} it was suggested that quantum nonlinearities might be related to Lorentz violation, leading one to
consider higher-energy tests. Although in this paper we initially adopt the more general possibility that $F$
might be nonlinear but Lorentz invariant, we find that Lorentz invariant nonlinearities are not likely to be
probed by neutrino oscillations. Furthermore, we leave open the possibility that $F$ might only be an effective
nonlinearity, summarising unknown microscopic physics, rather than a fundamental modification of quantum theory.

In our construction of nonlinear Dirac equations in \cite{NP1,np1a} we required the equation to be invariant
under the scaling $\psi \to \lambda \psi$, where $\lambda$ is an arbitrary constant, to ensure that the
wavefunction can be freely normalised as in the linear theory \cite{nonlin1,nonlin1a,nonlin1b}. This constraint
is desirable for fundamental theories and leads to nonpolynomial nonlinearities; however here we consider also
simpler polynomial nonlinearities (as effective theories) which do not have that invariance.

We restrict our study to the $\mu-\tau$ sector as this oscillation is more likely to be probed at higher
energies in the near future compared to the $e-\mu$ oscillation
\cite{future,fut1,fut2,fut3,fut4,fut5,fut6,fut7}. As in the standard formalism we take the weak eigenstates of
the neutrinos in the $\mu-\tau$ sector to be superpositions of the mass eigenstates,
    \begin{equation}\label{1.1}
        \psi_\alpha(x)=\sum_i U^*_{\alpha i}\psi_i(x)
    \end{equation}
where $\psi_\alpha(x)$ are the neutrino flavor eigenfunctions, $\psi_i(x)$ are the ``oscillating" eigenfunctions, and $U$ is the Leptonic mixing matrix.
For two-neutrino flavor oscillations the mixing matrix is
    \begin{equation}
        U=\left(\begin{array}{cc}\cos\theta&\sin\theta\\-\sin\theta&\cos\theta\end{array}\right).
\end{equation}
We assume the nonlinearity to be weak and so may approximate the mass eigenstates, to leading order
\cite{NP1,np1a}, by plane wave solutions with modified dispersion relations,
 \begin{equation}\label{1.4}
        \psi_i(\bfb{x}_i,t_i)= e^{-i(E_it_i-\bfb{p}_i\cdot\bfb{x}_i)} u_i(E,\bfb{p})
\end{equation}
where $u_i$ are spinors.

Thus after travelling a
distance $L$ between source and detector, and averaging over the unobserved travel time, as in \cite{Kayser},
one obtains the usual flavour change probability formula
\begin{equation}\label{1.11}
        P_{\nu_\alpha\rightarrow\nu_\beta}(E,L)=\sin^22\theta\sin^2(\frac{L\Delta p}{2})
    \end{equation}
 where $E$ is the beam energy, the momenta are taken in the direction of $L$ and $\Delta p = p_i -p_j$. Recall that in the $F=0$ case, with small masses and large energies, one has $\Delta p   =\Delta m^2 / 2E$ where $\Delta m^2 =m_i^2 -m_j^2$. Maximum oscillations occur when $L=L_0$, where
    \begin{equation}\label{1.12}
        \frac{L_0\Delta p}{2}=\frac{\pi}{2}\,\,\,\rightarrow\,\,\,L_0=\frac{\pi\hbar}{\Delta p} \, .
    \end{equation}
In the last step, we have restored the $\hbar$'s and $c$'s. $L_0$ is the oscillation length, the path-length needed for a neutrino of flavour $\alpha$ to maximally oscillate to a neutrino of flavour $\beta$. Thus the oscillation length in the conventional approach is given by
    \begin{equation}\label{1.13}
        L_0=\frac{2\pi \hbar cE}{\Delta m^2c^4} \, .
    \end{equation}

The expressions (\ref{1.11}) and (\ref{1.12}) are valid even in the nonlinear theory but with a modified $\Delta p$. Since we are adopting the plane wave approximation, the only difference from the conventional formalism will be modified dispersion relations in the expression for $\Delta p$.

We classify the types of $F$ in (\ref{nld}) into polynomial (P) or the non-polynomial (NP) forms studied in
\cite{NP1,np1a}, then into Lorentz violating (LV) or Lorentz invariant (LI). In \cite{NP1,np1a} we had studied
the NP type of $F's$ in a double expansion in the degree of nonlinearity $n$ and number of derivatives $d$, for
example
\begin{equation}
F=\left(\frac{\bar \psi\gamma^5\psi}{\bar \psi\psi}\right)^n
\end{equation}
has $d=0$ and degree $n$. In this paper, for simplicity we consider nonlinearities $F$ which consist of a single
term with $n=\alpha$ and structure $F=(*)^\alpha$ with $(*)$ containing at most one derivative, so that $F$ has
at most $\alpha$ derivatives. Furthermore we consider the general case of real $\alpha$, not necessarily an
integer. While non-integral powers might not be surprising in an effective theory, interestingly the specific
fraction $\alpha =1/3$ appears when one demands conformal invariance of a simple nonlinear Dirac equation
\cite{Russ,russa}. Also, as noted in \cite{NP1,np1a}, in the nonpolynomial case one can still preserve
separability for general $\alpha$ through an appropriate construction.

Since $F$ is a matrix in spinor space, we will consider two special cases, $F \propto I$ and $F \propto
\gamma^\mu$,  representing corrections to the mass or kinetic terms of the usual Dirac equation. As the
nonlinearity must be small on phenomenological grounds, we can compute the modified dispersion relations in
perturbation theory using the plane wave solutions of the linear theory, see \cite{NP1,np1a}.

We describe the Lorentz violating cases using background vector fields  $A_\mu$ as in \cite{coll,colla}. The
background fields may be interpreted as the effective coupling constants of an underlying microscopic theory
\cite{coll,colla}. In our case, {\it the background fields will simultaneously control the magnitude of both the
nonlinearity and the Lorentz violation.} A more detailed discussion of Lorentz violating nonlinear Dirac
equations is in \cite{NP1,np1a}.

The rest of the paper is structured as follows: In the next section we consider one example of a nonlinearity from the class NP-LV in detail and list results for other cases. In Section (3) we bound our parameters using current experimental data and then use our expressions for the modified oscillation probabilities to estimate corrections in future higher energy experiments. In Section (4) we discuss the other classes of nonlinearities while Section (5) concludes the paper.

The conventions we use are similar to those of the textbook \cite{IZ} and \cite{NP1,np1a}. We work in $3+1$
dimensional flat spacetime with metric $g^{\mu\nu}=(1,-1,-1,-1)$,
 set $\hbar=1=c$ and restore them as and when needed.  Note that while we work in flat space, the nonlinearity plausibly includes in an effective way the effects of gravity \cite{RP1}.

\section{The NP-LV class}

\subsection{$F_1$}
An example of an $F$ of the type considered in \cite{NP1,np1a} is given by
\begin{equation}
F_1=\left(A_\mu\frac{\bar \psi\gamma^\mu\psi}{\bar \psi\psi}\right)^\alpha \label{eg1}
\end{equation}
where $A_\mu$ is a real constant background field and $\alpha$ is any real number.
Thus this is a NP-LV type of nonlinearity of degree $n=\alpha$, with $F$ proportional to $I$ and no derivatives, $d=0$.

To be more specific, we suppose initially the background field to be given in the sun-centered celestial
equatorial frame \cite{frame}. As our expressions, such as (\ref{eg1}), are invariant under observer Lorentz
transformations \cite{coll,colla,NP1,np1a}, if need be, we can always express the inertial frame of the
earth-based (or space-based) experiments in terms of the sun-center celestial equatorial coordinates by
performing an observer Lorentz transformation \cite{frame}. However, as such a change of frame will not change
the order of magnitude of our quantities, the analysis we perform in Sect.(3) is essentially unaffected by a
switch of frames.

Perturbing around the plane wave limit gives $F_1=\left(\frac{A\cdot p}{m}\right)^\alpha$. The modified dispersion relation is then given by
\begin{equation}
p^2=m^2-2m\left(\frac{A\cdot p}{m}\right)^\alpha+O(A^{2\alpha}) \, .
\end{equation}
Expanding, we have
\begin{equation}
E^2-p^2=m^2-2m\left(\frac{A_0E-|\mathbf{A}||\mathbf{p}|\cos\phi}{m}\right)^\alpha
\end{equation}
where, $\phi$ is the angle between the spatial component of the background field and the neutrino momentum. For small masses
\begin{equation}
p\simeq E-\frac{m^2}{2E}+\left(A_0-|\mathbf{A}|\cos\phi\right)^\alpha\left(\frac{E}{m}\right)^{\alpha-1} \, . \label{a1}
\end{equation}
Assuming the temporal and the spatial components of the background field to be of similar order of magnitude, represented by $A$, we can rewrite the above as
\begin{equation}
p\simeq E-\frac{m^2}{2E}+A^\alpha\left(\frac{E}{m}\right)^{\alpha-1} \, . \label{a2}
\end{equation}
The momentum difference is given by
\begin{equation}
\Delta p=\frac{\Delta m^2}{2E}-E^{\alpha-1}\Delta\left(\frac{A^\alpha}{m^{\alpha-1}}\right)
\end{equation}
where $\Delta\left(\frac{A^\alpha}{m^{\alpha-1}}\right)=\frac{A_i^\alpha}{m_i^{\alpha-1}}-\frac{A_j^\alpha}{m_j^{\alpha-1}}$. Note that we have indicated a possible species dependence in the background gauge field. The oscillation length is
\begin{eqnarray}
L_0&=&\frac{2\pi E}{\Delta m^2-2E^{\alpha}\Delta\left(\frac{A^\alpha}{m^{\alpha-1}}\right)}
\end{eqnarray}
and may be written in the form
\begin{equation}
L_0=\frac{2\pi E}{\Delta m^2\left(1-X_1\right)} \label{xx}
\end{equation}

\subsection{Summary of Other NP-LV Cases}
We list here the other types of $F$ we study in this class.

\begin{itemize}
\item $F \propto I$, $n=d =\alpha $.
\begin{equation}
    F_{2}=\left[iA_\mu\left(\frac{\partial^\mu\bar \psi-\left(\partial^\mu\bar \psi\right)\psi}{2\bar \psi\psi}\right)\right]^\alpha
\end{equation}

\item {$F \propto \gamma^\mu$, $n=\alpha$, $d=0$}
\begin{equation}
    F_{3}=A_\mu\gamma^\mu\left[B_\nu\left(\frac{\bar \psi\gamma^\nu\psi}{\bar \psi\psi}\right)\right]^\alpha
\end{equation}

\item {$F \propto \gamma^\mu$, $n=d=\alpha$}
\begin{equation}
    F_{4}=A_\mu\gamma^\mu\left[iB_\nu\left(\frac{\bar \psi\partial^\nu\psi-\left(\partial^\nu\bar \psi\right)\psi}{2\bar \psi\psi}\right)\right]^\alpha
\end{equation}
\end{itemize}

All background fields are real. They are to be defined and approximated as in the previous subsection.
The dispersion relations are obtained by perturbing around plane waves as in the previous subsection to obtain the corresponding $X$'s in formula (\ref{xx}).

\section{Empirical Bounds and Estimates}
We have the modified oscillation lengths in the form
\begin{equation}
L_0=\frac{2\pi E}{\Delta m^2\left(1-X\right)}
\end{equation}
where $X$ is the leading order correction depending on the nonlinearity parameters and energy. Current neutrino
oscillation data fit the conventional neutrino mass scenario quite well. Still, as there are the usual
experimental uncertainties, we use those to estimate the size of $X$. From \cite{review}, we conservatively take
$X$ to be in the range $10\%$ to $0.01\%$ and will use this to constrain the range of values  which $\alpha$ can
take for each type of nonlinearity considered. Following that, we will estimate the value of $X$ in future
higher energy experiments.

Since the background fields play the dual role of Lorentz violating and nonlinearity parameters, we rewrite them as follows
\begin{eqnarray}
       \mbox{For $F_1$:}&A^\alpha&\rightarrow\,\,\,\epsilon_1\\
    \mbox{For $F_2$:}&A^\alpha&\rightarrow\,\,\,\epsilon_2\\
    \mbox{For $F_3$:}&AB^\alpha&\rightarrow\,\,\,\epsilon_3\\
    \mbox{For $F_4$:}&AB^\alpha&\rightarrow\,\,\,\epsilon_4
\end{eqnarray}
Upon restoring the factors of $\hbar$'s and $c$'s, the length dimensions of the parameters is given by
\begin{eqnarray}
    \left[\frac{\epsilon_1}{\hbar}\right]&=&L^{-1}\\
    \left[\frac{\epsilon_2}{\hbar}\right]&=&L^{\alpha-1}\\
    \left[\frac{\epsilon_3}{\hbar}\right]&=&L^{-1}\\
    \left[\frac{\epsilon_4}{\hbar}\right]&=&L^{\alpha-1}
\end{eqnarray}

\subsection{Current Experiments}
The natural length scale in the linear theory is of course the Compton wavelength, $\lambda_c=\frac{h}{mc}$, and by using this one may express the nonlinear equation in dimensionlesss form by introducing small dimensionless parameters that characterize the nonlinearity/Lorentz violation. From current data, we may take the size of the small dimensionless LV parameters to be
\begin{eqnarray}
f&\sim& 10^{-27}
\end{eqnarray}
where $f$ is a dimensionless parameter associated with Lorentz violation \cite{LV}. This parameter $f$ is
implicit in the nonlinear parameter $\epsilon$'s and will be factored out below in Eq.(\ref{fff}). If the
parameters are neutrino species dependent then we assume $\Delta f \sim f$. The relevant neutrino data are taken
to be
\begin{eqnarray}
\Delta m^2&=&2.5\times 10^{-3} eV^2\\
E&=& 100GeV \, ,
\end{eqnarray}
that is we use the mean beam energy $E=100 GeV$. Since we are only interested in order of magnitude estimation,
it is sufficient to consider only a single energy. We assume that the mass of neutrinos is of the same order of
magnitude as $\sqrt{\Delta m^2} $. In the same, ``order of magnitude" spirit, we estimate the expression
$\Delta(m\epsilon)$ by $\left(\Delta m\right)\left(\epsilon\right)$.

The nonlinear parameter itself may be written, taking $F_2$ as an example,
        \begin{equation}
            \epsilon_2 = \lambda^{\alpha-1} f \label{fff}
        \end{equation}
where $\lambda$ is the characteristic length scale of the linear theory.

As the actual mass of neutrinos may be 1 or 2 orders of magnitude larger than what we have assumed above, we compensate for this possibility and the other approximations above by taking
        \begin{eqnarray}
            X&&\mbox{ of order $10^{-1}$ to $10^{-4}$}
        \end{eqnarray}

The various expressions of $X$'s upon restoring the $\hbar$'s and $c$'s, and using the above assumptions, are given by
\begin{eqnarray}
    X_1&=&\frac{2\hbar cf}{\lambda\left(\Delta m^2c^4\right)^{(\alpha_1+1)/2}}E^{\alpha_1} \label{x1}\\
    X_2&=&\frac{2f\lambda^{\alpha_2-1}}{(\hbar c)^{\alpha_2-1}\left(\Delta m^2c^4\right)^{1/2}}E^{\alpha_2} \\
    X_3&=&\frac{2\hbar cf}{\lambda\left(\Delta m^2c^4\right)^{(\alpha_3+2)/2}}E^{\alpha_3+1} \\
    X_4&=&\frac{2f\lambda^{\alpha_4-1}}{(\hbar c)^{\alpha_4-1}\Delta m^2c^4}E^{\alpha_4+1}  \label{x4}
\end{eqnarray}

Our procedure is as follows: Since $X$ depends on the characteristic length, $\lambda$, we invert the
relationship to plot $\lambda$ as a function of $\alpha$ using the values for $f,m,E$  mentioned above and for
the chosen range of $X$ values's. From these plots, we can determine the values of $\alpha$'s for which  the
characteristic length scale $\lambda$ lies within the range $10\lambda_c$ to $0.1\lambda_c$\footnote{Note that
we have used the similar argument as above for Compton wavelength. That is Compton wavelength is given by
$\lambda_c=hc/\sqrt{\Delta m^2c^4}$}. The plots of $\lambda$ versus $\alpha$ for $X_1$ is shown in Figure 1.
From Figure 1, and similar ones for the other $X$'s, we obtain the corresponding range of values for the
$\alpha$'s:
\begin{eqnarray}
\mbox{For $F_1$}&&\mathbf{1.8}\leq\alpha_1\leq2.2\\
\mbox{For $F_2$}&&\mathbf{1.7}\leq\alpha_2\leq2.1\\
\mbox{For $F_3$}&&\mathbf{0.8}\leq\alpha_3\leq1.2\\
\mbox{For $F_4$}&&\mathbf{0.8}\leq\alpha_4\leq1.1
\end{eqnarray}
Since we expect the nonlinear effects to be small, the $\alpha$'s are more likely to be observed near the lower bound in $X$. These $\alpha$'s are indicated in boldface in the above equations. Note that the bounds depend on the choice of $X$ and $\lambda_c$ and so must be updated as more accurate data becomes available.

\subsection{Future Experiments}
We now reverse the argument. We plot $X$ versus $E$ for the range of $\alpha$ values determined in the previous
section. From these plots, we can estimate the values of $X$'s expected in  future experiments where higher
energies will be available \cite{future,fut1,fut2,fut3,fut4,fut5,fut6,fut7}.

The plots are shown in Figures 2 to 4 for some cases and parameter values.
The general trend of $X(E)$ can be seen from equations (\ref{x1} -\ref{x4}). Of course since we evaluated the $X$'s perturbatively, the expressions and plots are valid only for small $X$, while for larger $X$ the indicated trend is only qualitative.

\section{Other Classes of Nonlinearities}

\subsection{P-LV}
Polynomial type of nonlinearities lead to a non-separable Dirac equation.
For example from the NP-LV, $F \propto I$  cases considered before one may remove the denominators to get corresponding P-LV type of nonlinearities. But now one sees that the modified dispersion relation will depend on the normalisation chosen for the wavefunctions. If one chooses the usual plane wave normalisation such that $\bar{\psi} \psi =1$ then the results for the P-LV cases mentioned above would be the same as for the NP class.

So to obtain new results let us explore the popular normalisation $\psi^{\dag} \psi=1$ which makes
$\bar{\psi} \psi = m/E$. But this energy factor from the normalisation will cancel that from the nonlinearity in the two $F \propto I$ P-LV cases obtained in the previous paragraph, thus making the modified dispersion relation energy independent. Therefore only the $F \propto \gamma^\mu$ cases are relevant. We label these as
\begin{eqnarray}
F_5 &=& A_\mu \gamma^\mu (\bar{\psi} \psi)^\alpha \\
F_{6} &=& A_\mu\gamma^\mu \left[ iB_\nu \left(\bar \psi\partial^\nu\psi-\left(\partial^\nu\bar \psi\right)\psi\right)\right]^\alpha
\end{eqnarray}
Note that while $F_6$ is just the numerator of $F_4$, $F_5$ is not the numerator of $F_3$ as now we have a simpler way to generate a P-LV $n=\alpha, d=0$ nonlinearity. The corresponding $X$'s are
\begin{eqnarray}
X_5 &=& \frac{2\hbar c\left(\Delta m^2c^4\right)^{(\alpha-2)/2}\Delta f}{\lambda}E^{1-\alpha}\\
X_6 &=&\frac{2^{\alpha+1}\lambda^{\alpha-1}\left(\Delta m^2c^4\right)^{(\alpha-2)/2}\Delta f}{\left(\hbar c\right)^{\alpha-1}}E
\end{eqnarray}
and the range of $\alpha$ values are found to be

\begin{eqnarray}
 -1.2< \alpha_5 \le\mathbf{-0.8} \\
\mathbf{5.8}\le \alpha_6 \le 133
\end{eqnarray}
Again the $\alpha$'s in boldface are near the more likely values. We see that for $F_5$, we need negative values of $\alpha$. This would make nonlinear extensions like $F_5$ non-local; such non-localities might arise as effective corrections arising from some more basic quantum field theory. For $F_6$, we see that $\alpha=133$ implies a huge number of derivatives for $X=10\%$ and it becomes a more reasonable $\alpha=5.8$ for smaller $X$.

\subsection{NP-LI}
If the nonlinearity is Lorentz invariant, the modified dispersion relations remain covariant $E^2 =p^2 + M^2$
but with an effective mass $M$ that depends on the nonlinearity parameters \cite{NP1,np1a}. If we take these
nonlinearity parameters to be non-universal, meaning that the different neutrinos get different effective mass
corrections, then the oscillation probabilities are modified. However as the NP type of nonlinearities are
invariant to the normalisation of $\psi$, the modification is energy independent and thus not relevant for
high-energy tests.

\subsection{P-LI}
 Choosing the same energy dependent normalisation as in the P-LV cases discussed above, and noting the discussion in the previous subsection, one sees that in P-LI cases such as $(\bar\psi \psi)^\alpha$ the $X$ decreases with increasing energy and thus becomes irrelevant at high energies. (We have looked at a few simple cases of $F$ in the class P-LI and they show a similar trend).

\section{Discussion}

The way we have parametrized our corrections, the modification to the conventional neutrino oscillation probabilities  may be described in terms of an effective energy dependent mass-squared difference,
\begin{equation}
\Delta m^2_{eff}(E) = (1-X) \Delta m^2 \label{eff}
\end{equation}
Hopefully such an effect may be detectable with higher statistics and energies in future experiments. Note that our $X$'s are positive because we took the constant in front of the $F$'s in (\ref{nld}) to be positive and absorbed it into the nonlinearity parameter. More generally then, the right-hand-side of  (\ref{eff}) should read $(1 \pm X) \Delta m^2$ so that the effective mass might increase or decrease with energy.

Among the various types of nonlinearities we have studied,  we find that only six have the potential to be detected in future higher energy experiments through their increasing energy dependent effect on the neutrino oscillation probabilities. In the two $F \propto I$ cases, $F_1, F_2$, each of the discrete symmetries is preserved while in the remaining
 $F \propto \gamma^\mu$ cases the nonlinearities are $C$ and $CPT$ odd. Thus the discrete symmetries might be one way of partly discriminating among the possibilities. Some interesting behaviour is seen for $F_5,F_6$ but note that in those cases the nonlinearity is dependent on the energy dependent normalisation.

Since we worked in the plane wave approximation, the above-mentioned effects only probe modified dispersion
relations rather than the nonlinearity itself. However there are various ways of distinguishing our results from
other proposals in the literature. Firstly, we found that in modelling the nonlinearity by a single term $F \sim
(*)^\alpha$ in the evolution equation, $\alpha$ turns out to have generically noninteger values so that
\begin{equation}
{ \Delta m^2 (E) -\Delta m^2(0) \over \Delta m^2(0) } \propto \pm E^{\beta}
\end{equation}
for some positive and typically fractional $\beta$. While even this can be obtained simply from a modified
dispersion relation, \cite{Joy,gasperini,alfaro,adunas,ram,morgan}, independent of a quantum nonlinearity, the
availability of a quantum evolution equation in our approach will enable a more refined analysis of the
phenomena when sufficient data become available. We also emphasize that our nonlinearity simultaneously violates
Lorentz invariance and that is an additional distinguishing feature.

Further work in this direction would involve going beyond the plane wave approximation, leading to genuine nonlinear effects and perhaps leading to an understanding of the mixing angles too \cite{RP1}. Also, a subleading  directional dependence of the oscillation probability can be examined by using (\ref{a1}) instead of the approximation (\ref{a2}). Finally, one should explore if current oscillation data can be fit using purely energy dependent effective neutrino masses as suggested, for example, in \cite{RP1}.

\newpage

\section{Figures}

\noindent Figure 1: This is a plot of $\lambda$ vs $\alpha$ for $X_1$. The vertical axis, plotted in log scale, has units of metre while the horizontal axis is dimensionless. The solid and the dashed lines represent $X=10^{-1}$ and $X=10^{-4}$ respectively. The horizontal lines are the bounds $10\lambda_c$ and $0.1\lambda_c$.
\bigskip

\noindent Figure 2: This is the log-log plot of $X_1$ vs energy. The full and dashed lines have $\alpha$ values of 1.8 and 2.2 respectively. Here we have set $\lambda=\lambda_c$. Note that $X_3(E)$ has an identical plot to Fig.2 after the following redefinition of $\alpha$: The full and dashed lines have $\alpha$ values of 0.8 and 1.2 respectively, while $\lambda=\lambda_c$.
\bigskip

\noindent Figure 3: This is the log-log plot of $X_2$ vs energy. The full and dashed lines have $\alpha$ values of 1.7 and 2.1 respectively; $\lambda=\lambda_c$.
\bigskip

\noindent Figure 4: This is the log-log plot of $X_4$ vs energy. The full and dashed lines have  $\alpha$ values of 0.8 and 1.1 respectively; $\lambda=\lambda_c$.

\newpage

\begin{figure}[h!]
\includegraphics[width=13cm]{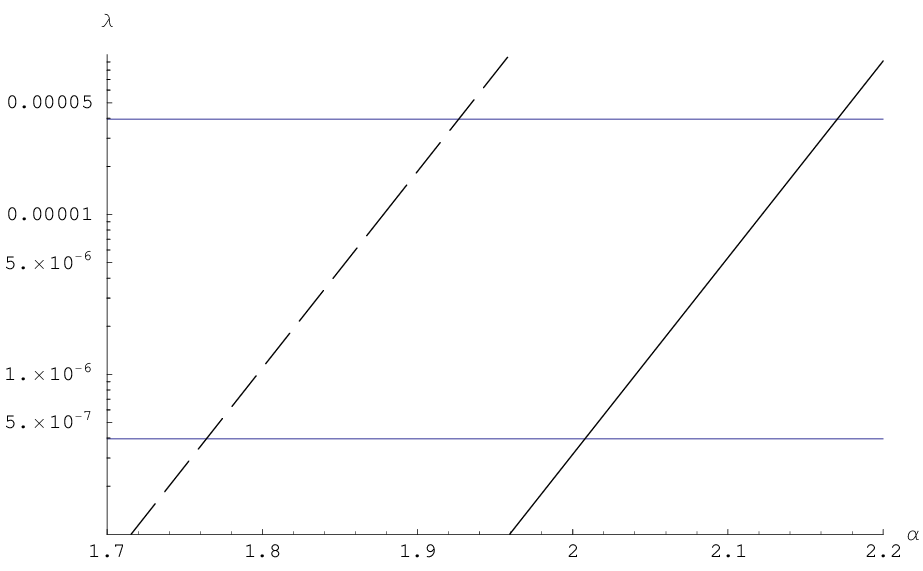}
\caption{}
\end{figure}

\begin{figure}[h!]
\includegraphics[width=13cm]{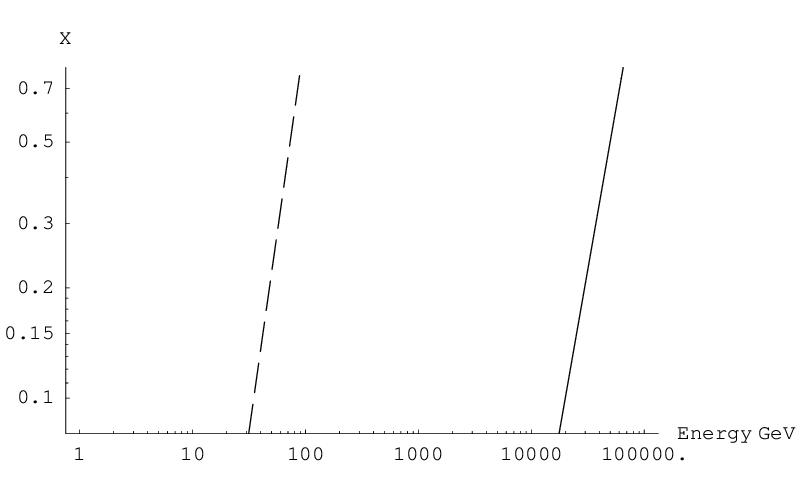}
\caption{}
\end{figure}

\begin{figure}[h!]
\includegraphics[width=13cm]{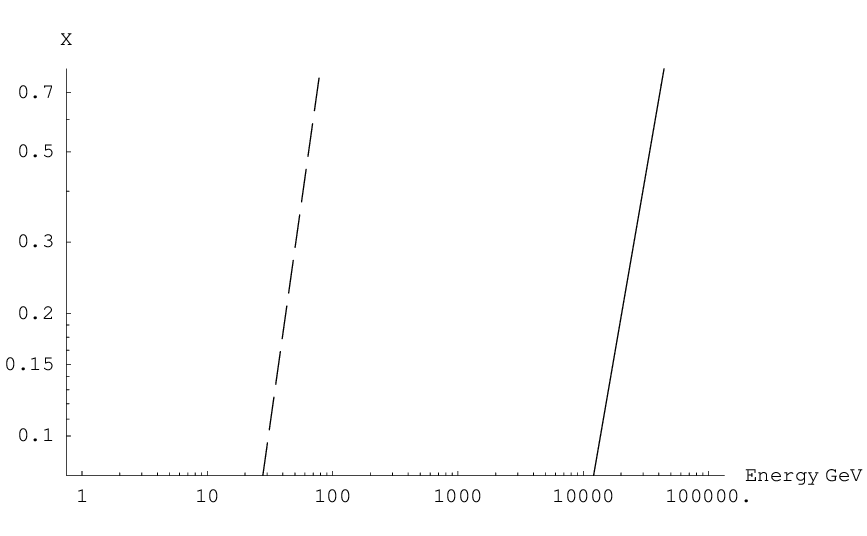}
\caption{}
\end{figure}

\begin{figure}[h!]
\includegraphics[width=13cm]{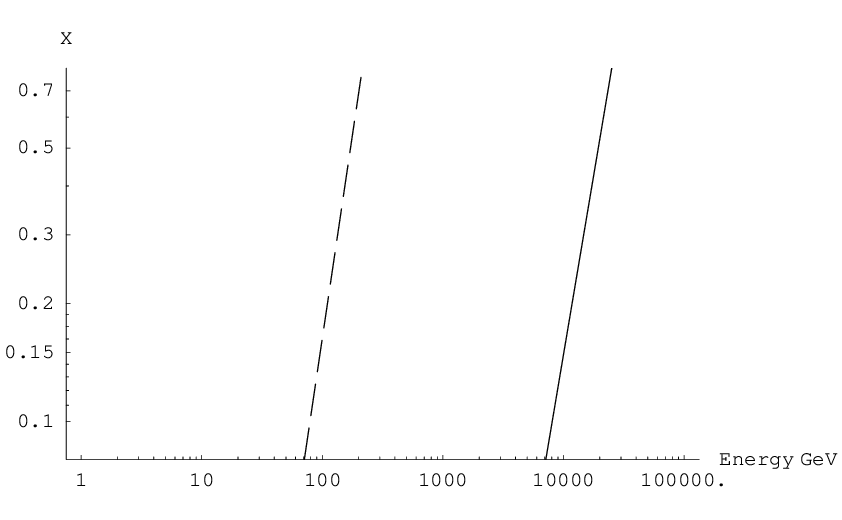}
\caption{}
\end{figure}

\end{document}